\documentclass[useAMS,usenatbib,usegraphicx]{mn2e}

\title[Helioseismic Measurements in the Solar Envelope]
      {Helioseismic Measurements in the Solar Envelope Using Group Velocities of Surface Waves}
\author[S. V. Vorontsov, V. A. Baturin, S. V. Ayukov, V. K. Gryaznov]
 {S. V. Vorontsov$^{1,2}$, V. A. Baturin$^3$, S. V. Ayukov$^3$, V. K. Gryaznov$^4$ \thanks{E-mail: S.V.Vorontsov@qmul.ac.uk}\\
$^{1}$Astronomy Unit, Queen Mary, University of London, Mile End Road, London E1 4NS, UK\\
$^{2}$Institute of Physics of the Earth, B.Gruzinskaya 10, Moscow 123995, Russia\\
$^{3}$Sternberg State Astronomical Institute of Lomonosov Moscow State University,
      Universitetsky prospect 13, Moscow 119992, Russia\\
$^{4}$Institute of Problems of Chemical Physics RAS, Academician Semenov avenue 1,
      Chernogolovka, Moscow region 142432, Russia}
\begin{document}

\date{Accepted 2012 December 00. Received 2012 December 00; in original form 2012 December 00}

\pagerange{\pageref{firstpage}--\pageref{lastpage}} \pubyear{2006}

\maketitle

\label{firstpage}

\begin{abstract}

At intermediate and high degree $l$, solar p- and f modes can be considered as surface waves.
Using variational principle, we derive an integral expression for the group velocities of the surface waves
in terms of adiabatic eigenfunctions of normal modes, and address the benefits of using group-velocity measurements
as a supplementary diagnostic tool in solar seismology.
The principal advantage of using group velocities, when compared with direct analysis of the oscillation frequencies,
comes from their smaller sensitivity to the uncertainties in the near-photospheric layers.
We address some numerical examples where group velocities are used to reveal inconsistencies between the solar models and the seismic data. Further, we implement the group-velocity measurements to the calibration of the specific entropy,
helium abundance Y and heavy-element abundance Z in the adiabatically-stratified part of the solar convective envelope,
using different recent versions of the equation of state. The results are in close agreement with our earlier measurements
based on more sophisticated analysis of the solar oscillation frequencies (Vorontsov et al. 2013, MNRAS 430, 1636).
These results bring further support to the downward revision of the solar heavy-element abundances
in recent spectroscopic measurements.
\end{abstract}


\section{Introduction}
The major difficulty in the seismic measurements of the solar internal structure
comes from the uncertain effects of the outermost solar layers (the photosphere and layers immediately below),
where trapped acoustic waves are reflected to the solar interior.
The difficulty originates from both the uncertainties in the theoretical modeling of these layers
(e. g. effects of the penetrative convection), and from poor understanding of the physics of wave propagation there
(non-adiabatic effects). This difficulty is behind the dominant source of mismatch between the observational and theoretical
p-mode frequencies. The discrepancy increases with frequency (as the upper turning points of the p modes move upwards), reaching
values of the order of one percent at frequencies of maximum oscillation power (about 3 mHz). In standards of solar seismology,
one percent is a huge quantity: the frequencies of solar oscillations are measured with precision better than one part in $10^4$,
and it is this high precision which enables the seismic data with its unique diagnostic capability.

To allow an accurate diagnostic of the deep interior, the near-surface uncertainties are suppressed,
in one way or another, by a proper design of helioseismic inversion technique. Principally, the separation of the uncertain effects
is made possible by the relatively small values of the sound speed in the subsurface layers, which makes the acoustic ray paths
nearly vertical there, when the degree $l$ of the oscillations is not too high. The possibility of separating the uncertain effects
is most transparent when high-frequency asymptotic analysis is implemented to describe the solar p modes: the subsurface effects
bring a frequency-dependent phase shift $\alpha(\omega)$ of the standing acoustic wave, which does not depend on the degree $l$ when $l$ is small.

Separation of the near-surface uncertainties comes for a price of loosing valuable diagnostic information.
An example is He II ionization region, the domain which is particularly important for measuring the solar abundances and for
the calibration of the equation of state. For p modes of low degree $l$, the signal of He ionization is also seen as a frequency-dependent
``surface phase shift''; this signal is suppressed together with near-surface uncertainties when an arbitrary function of frequency is allowed for $\alpha(\omega)$. At higher degree $l$, we meet another difficulty. In high-precision measurements, the acoustic
waves in the subsurface layers can no longer be considered as purely vertical, and at least a first-order correction shall be added
to $\alpha(\omega)$ to account for the resulted dependence of the surface phase shift on the degree $l$, as another function of frequency
multiplied by $l(l+1)$ \citep{Brodsky93}. But according to the asymptotic description, allowing the degree dependence to the surface term
widens the family of possible solutions in the deep interior \citep{Gough95}.

It is desirable, therefore, to extend the set of diagnostic tools, implemented in solar seismology for analyzing the oscillation frequencies, by adding new tools which respond differently to the near-surface uncertainties, and suppress these uncertainties
in a different way. This will allow more options for the cross-validation of the results, to make them more reliable.
An issue of particular importance is possible effects of systematic errors in frequency measurements; these errors may be significantly
bigger than the reported observational uncertainties \citep[see e. g.][which we refer below as Paper I]{Vorontsov13a}.
In general, systematic errors propagate differently to the results when different techniques of data analysis are implemented, and
using different tools brings better chances to detect these errors.

In this paper, we consider solar oscillations of intermediate and high degree $l$ as surface waves, and address the diagnostic properties of group velocities of these waves. The concept of group velocity is known to be a valuable tool in terrestrial seismology, where it is
applied to study the propagation of Love's and Rayleigh's waves \citep[see e. g.][]{Dahlen98}. Section 2 contains a general discussion, based on the integral representation of the group velocity in terms of adiabatic eigenfunctions of normal modes, developed in the Appendix.
In section 3, we test the diagnostic potential of group velocities by addressing the agreement of several solar models
with observational data. In section 4, we implement the group-velocity analysis to the calibration of the main parameters
of the solar convective envelope: specific entropy in the adiabatically-stratified layers, helium abundance $Y$ and heavy-element
abundance $Z$. Section 5 contains a short discussion.

\section[]{The group velocity}
With temporal dependence separated as $\exp(-i\omega t)$, the displacement field of the oscillations specified by a particular spherical harmonic is
\begin{equation}
{\bf u}=\hat{\bf r}U(r)Y_{lm}(\theta,\phi)+V(r)\nabla_1Y_{lm}(\theta,\phi),
\end{equation}
where $\nabla_1$ is horizontal component of the gradient operator, $\nabla_1=\hat{\bf\theta}\partial/\partial\theta+\hat{\bf\phi}\sin^{-1}\theta\,\partial/\partial\phi$, and unit vectors are designated by hats. The horizontal wavenumber at the photospheric level is $k_H=L/R_\odot$, with $L^2=l(l+1)$. The horizontal phase velocity $v_p$ and group velocity $v_g$ are
\begin{equation}
v_p={\omega\over L}R_\odot,\quad\quad
 v_g=\left({\partial\omega\over\partial L}\right)_nR_\odot,
\end{equation}
where the derivative is taken at constant radial order $n$.

Using self-adjoint properties of the equations of linear adiabatic
oscillations, it is shown in the Appendix that
\begin{equation}
{v_g\over v_p}\equiv
 \left({\partial\ln\omega\over\partial\ln L}\right)_n=
 {\int\limits_0^R\rho_0r^2L^2V^2dr
 +{L^2\over 4\pi G\omega^2}\int\limits_0^\infty P^2dr
 \over
 \int\limits_0^R\rho_0r^2\left(U^2+L^2V^2\right)\,dr}\,,
 \end{equation}
where $P=P(r)$ describes the Eulirean perturbation $\psi'$ to the gravitational potential as
\begin{equation}
\psi'=-P(r)Y_{lm}(\theta,\phi).
\end{equation}
At high degree $l$, the effects of gravity perturbation are small, and the second term in the nominator of the equation (3)
can be neglected. The first term in the nominator is proportional to the mean kinetic energy of the horizontal motions;
the denominator is proportional to the total kinetic energy. When the effects of gravity perturbations are small, the ratio $v_g/v_p$ is thus the ratio of the horizontal kinetic energy to the total kinetic energy.

An important property of the group velocity (when compared with phase velocity) is its enhanced sensitivity to the stratification
of the inner part of the acoustic cavity, where the horizontal kinetic energy is localized (Fig. 1). Closer to the surface, the acoustic ray paths become nearly vertical, and the kinetic energy is dominated by the vertical motion.
\begin{figure}
\begin{center}
\includegraphics[width=1.0\linewidth]{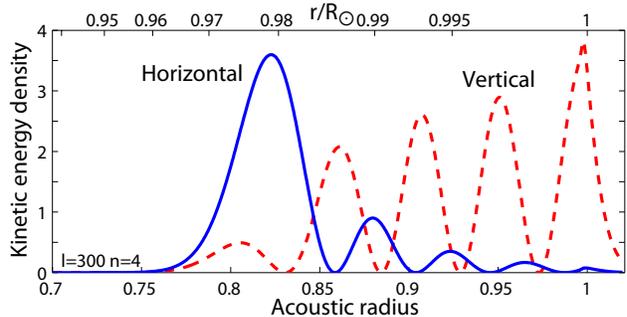}
\end{center}
\caption{Horizontal and vertical kinetic energy densities for $p_4$ mode of $l=300$. The energy per unit depth
in the integrals in equation (3) was multiplied with adiabatic sound speed $c(r)$ to account for the rescaling
of the independent variable from geometrical to acoustic radius.}
\label{f1}
\end{figure}

The concept of group velocity is not restricted to high-degree modes; it extends formally to all the non-radial modes
when we consider the degree $l$ as a continuous parameter. The ratio $v_g/v_p$ in the degree range $0\le l\le 300$,
calculated for the reference solar model S of \citet{Christensen96}, is shown in Fig.2.
\begin{figure}
\begin{center}
\includegraphics[width=1.0\linewidth]{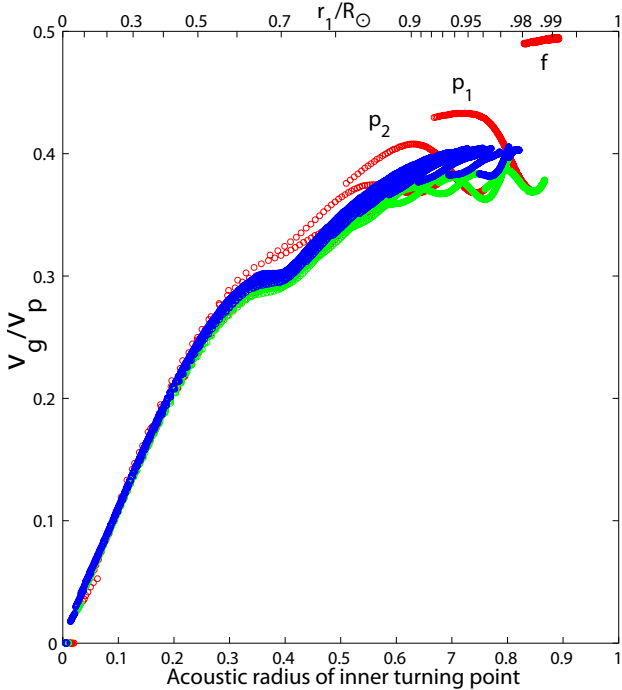}
\end{center}
\caption{Ratio of group and phase velocities calculated for the solar model
 in the degree range $0\le l\le 300$ and frequency range
 1mHz$\le\omega/2\pi\le$5mHz. Red circles show modes with frequencies below 2mHz,
 green---between 2 and 3mHz, and blue---above 3mHz. Upper scale is the position
 of the inner turning point in radius.}
\label{f2}
\end{figure}
For modes penetrating deep into the solar interior, the ratio
$v_g/v_p$ tends to collapse to a single function of the penetration
depth (Fig. 2), the behaviour which reflects the
high-frequency asymptotic properties of solar p modes. The rapid
variation seen at $r_1\approx 0.7R_\odot$ comes from the rapid
change in the sound-speed gradient at the base of the convection
zone. The prominent fluctuations in $c_g/c_p$ exhibited by modes
with turning points closer to the surface are produced by the
rapid variation of the adiabatic exponent in the He ionization
region.

For f modes, the ratio $v_g/v_p$ is close to 1/2, which
reflects equipartition of the kinetic energy between horizontal
and vertical motions. Indeed, using an approximate dispersion relation of high-degree solar f modes as
$\omega^2\simeq k_Hg_0(R_\odot)$, where $g_0(R_\odot)$ is surface gravity, we have
$(\partial\ln\omega/\partial\ln L)_{n=0}\simeq 1/2$.

Introducing the derivative $(\partial\omega/\partial L)_n$, we extend the family of solutions to the oscillation equations
from integer to continuous values of degree $l$. This extension can be achieved simply by allowing $l$ to take arbitrary values in the ordinary differential equations resulted from variable separation; it can be viewed as a result of relaxing the periodic
boundary conditions in angular coordinates.

We now proceed with similar generalization but allowing non-integer values to the radial order $n$. This is equivalent to allowing
a continuous variation to the radial phase integral, which takes values of $\pi(n+1)$ at resonant frequencies.
The only limitation here is that the phase has a well-defined measure only for functions with nearly-harmonic behaviour.
Such a behaviour is exhibited by the solutions to the oscillation equations in the adiabatically-stratified part of the solar
convective envelope, owing to their high-frequency asymptotic properties (the wave propagation here is close to that of purely
acoustic waves). Specifically, the near-harmonic behavior is exhibited with best accuracy by eigenfunction
$\psi_p(\tau)$ defined as \citep[see][and Paper I]{Vorontsov91b}
\begin{equation}
\psi_p=\rho_0^{-1/2}r\left({1\over c^2}-{\tilde w^2\over r^2}\right)^{1/4}
  \left(1-{N^2\over\omega^2}\right)^{-1/2}p_1,
\end{equation}
where $p_1=p_1(r)$ describes the Eulerian pressure perturbations $p'$ as
\begin{equation}
p'=p_1(r)Y_{lm}(\theta,\phi),
\end{equation}
\begin{equation}
\tilde w^2={L^2\over\omega^2},
\end{equation}
(this parameter specifies the radial position $r_1$ of the inner turning points), and independent variable $\tau$ satisfies
\begin{equation}
{d\tau\over dr}={1\over c}\left(1-\tilde w^2{c^2\over r^2}\right)^{1/2}.
\end{equation}
Two linearly-independent solutions to the oscillation equations in Cowling approximation (which is applicable
in the low-density envelope) are $\psi\simeq\sin(\omega\tau)$ and $\psi\simeq\cos(\omega\tau)$.

Using variational principle for evaluating the variation with frequency of phases of the inner and of the outer solutions
(which satisfy inner and outer boundary conditions, respectively) at a boundary taken in the domain where asymptotic description is
applicable, we obtain (see Appendix)
\begin{equation}
\left({\partial\omega\over\partial n}\right)_L
={\pi\over 2\omega^2}{\psi_p^2+{1\over\omega^2}\left({d\psi_p\over d\tau}\right)^2
\over\int\limits_0^R\rho_0r^2\left(U^2+L^2V^2\right)dr}.
\end{equation}
In analogy with $(\partial\omega/\partial L)_n$, considered as an angular component of the group velocity,
$(\partial\omega/\partial n)_L$ can be considered as a mean group velocity in radial direction.

The dependence of both the $(\partial\omega/\partial L)_n$ and $(\partial\omega/\partial n)_L$ on the near-surface uncertainties
comes principally from the dependence on these uncertainties of the denominator in the expressions (3) and (9), which is the same
(mode's kinetic energy). This observation suggests using the ratio
\begin{equation}
\gamma(L,n)=\left({\partial\omega\over\partial L}\right)_n\Big/\left({\partial\omega\over\partial n}\right)_L
\end{equation}
as diagnostic quantity, which will allow to suppress the effects of the near-surface uncertainties.
In the simplest way, this quantity can be evaluated from the mode frequencies using central differences as
\begin{equation}
\gamma_{ln}={\omega_{l+1,n}-\omega_{l-1,n}\over\omega_{l,n+1}-\omega_{l,n-1}}.
\end{equation}
According to the equation (9), the inverse of $(\partial\omega/\partial n)_L$ can be considered as ``mode mass''.
The difference with traditional definition of the mode mass (mode energy at unit surface amplitude) is that the surface
amplitude is replaced by the amplitude in the propagation domain. The principal advantage is that with this definition,
the ``mode mass'' becomes an observable quantity.

The diagnostic properties of $\gamma(L,n)$ defined by the equation (10) can be seen better if we extend the comparison
with high-frequency asymptotic analysis somewhat further. In the leading-order approximation, the asymptotic eigenfrequency equation is
\begin{equation}
\omega F(\tilde w)=\pi\left[n+\alpha(\omega)\right],
\end{equation}
where
\begin{equation}
F(\tilde w)=\int\limits_{r_1}^R\left(1-\tilde w^2{c^2\over r^2}\right)^{1/2}{dr\over c}.
\end{equation}
The left-hand side of the equation (12) is the radial phase integral $\int_{r_1}^Rk_rdr$ of a purely acoustic wave,
$\alpha(\omega)$ is the frequency-dependent ``surface phase shift''. The sound-speed profile $c(r)$ can be recovered from
$dF(\tilde w)/d\tilde w$ using Abel's integral transform applied to the equation (13).

Differentiating both sides of the equation (12) in frequency $\omega$, first at $L$=const, then at $n$=const, and subtracting
the results, we have
\begin{equation}
{dF\over d\tilde w}=-\pi\left({\partial\omega\over\partial L}\right)_n\Big/\left({\partial\omega\over\partial n}\right)_L,
\end{equation}
and we see that the influence of the unknown behaviour of $\alpha(\omega)$ on the results of the sound-speed inversion
is successfully eliminated \citep[it is indeed the way in which some first helioseismic sound-speed inversions were performed, see e. g.]
[]{Vorontsov89}. Comparing the equations (10) and (14), we see that $\gamma(L,n)$ is expected to be largely
insensitive to the near-surface uncertainties.

\section[]{Testing solar models with seismic data}
In this section, we test the ability of group-velocity measurements to reveal inconsistencies between the solar models
and the observational data. In these tests, we compare the $\gamma_{ln}$-values measured from the solar oscillation frequencies
with those obtained from the eigenfrequencies of solar models.

Fig. 3(a) shows the difference in $\gamma_{ln}$ between the Sun and the reference model S of \citet{Christensen96}. The observational
frequencies were obtained by averaging the results of 15 years of SOHO MDI measurements (this observational data set is discussed in
more detail in Paper I, where it is designated as data set 1). A prominent mismatch is seen in the group-velocity data for waves with turning points just below the convective envelope, which indicates an inadequate description of the seismic stratification
in the solar tachocline.
\begin{figure}
\begin{center}
\includegraphics[width=1.0\linewidth]{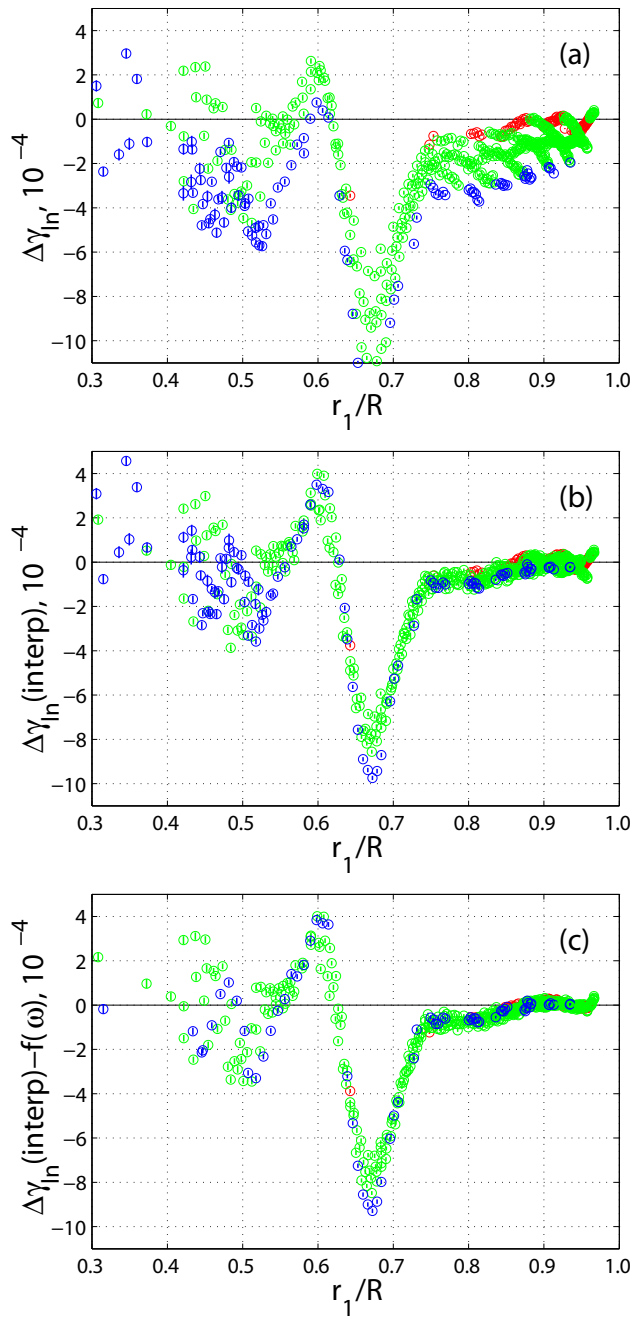}
\end{center}
\caption{(a) difference between observational values of $\gamma_{ln}$ and those of the reference model S.
(b) same residuals but obtained with interpolated values of $\gamma_{ln}$ (see text)
(c) as (b), but after a slowly-varying function of frequency was subtracted from the residuals.
Red circles show the results obtained at frequencies below 2 mHz, green circles---between 2 and 3 mHz,
and blue circles---with data above 3 mHz.}
\label{f3}
\end{figure}

For waves confined in the convective envelope, the mismatch is moderately small at low frequencies (below 2 mHz),
but grows significantly when frequency increases. This behaviour is induced by the systematic difference
between observational and theoretical frequencies, which grows when frequency increases. As a result, the $\gamma_{ln}$-values
evaluated with using the equation (11) from observational and theoretical frequencies correspond to surface waves with different
penetration depth. A simple way to reduce this effect is to replace the theoretical values of $\gamma_{ln}$
with values obtained by the interpolation along the p-mode ridge to proper values of $\tilde w$ (defined by equation 6).
The result is shown in Fig. 3(b); as expected, the residuals now fall much closer to a single
function of the penetration depth. Small, but systematic fluctuations around the common trend remain in the residuals even after the
interpolation. These are due to the fact that that the observational and theoretical values of $\gamma$, reduced to the same
penetration depth, are still measured at different frequencies (as a result, the phase of the wave function
in the He II ionization region is distorted). We can make the signal which brings information about inconsistencies in deep interior
cleaner still by subtracting a common function of frequency: the result is shown in Fig. 3(c) (in this computation, $f(\omega)$ was
obtained by approximating the residuals in the domain $r_1>0.85 R_\odot$ by polynomial of degree 10 in frequency $\omega$).

The model was than corrected by helioseismic inversion to bring it into agreement with seismic data. The resulted correction to the sound-speed profile is shown in Fig. 4(a). The inversion technique is described in Paper I; it results in another hydrostatic
model, which allows a new set of eigenfrequencies to be calculated. We processed these frequencies in the same way as discussed above,
to test the new model against the solar data. This is an important test: the solution in the deep interior may not be unique
because two arbitrary functions of frequency were allowed by the inversion to account for near-surface effects.
The result is shown in Fig. 4(b): there is no signal in the residuals. As the solar data were processed by the inversion in a very different manner, we now have better confidence in the results.
\begin{figure}
\begin{center}
\includegraphics[width=1.0\linewidth]{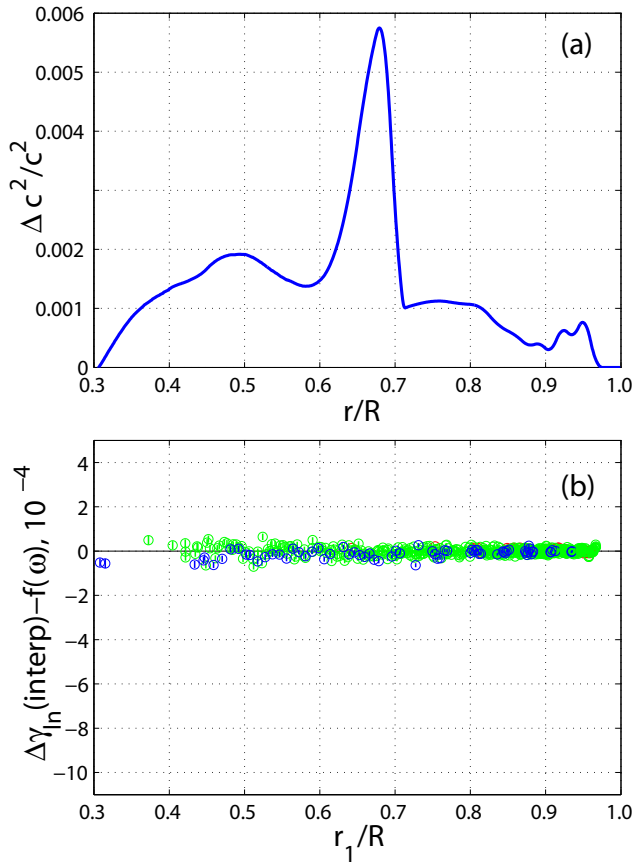}
\end{center}
\caption{(a) difference in the sound speed between the Sun and the reference model S, obtained in helioseismic structural inversion
with observational frequencies. (b) As Fig. 3(c), but obtained with eigenfrequencies of the model resulted from the inversion.}
\label{f4}
\end{figure}

Fig. 5 shows the results of similar tests but performed with two solar models having nearly-optimal parameters of the
adiabatically-stratified part of the convective envelope (specific entropy and chemical composition) which were measured by the seismic
calibration described in Paper I.  Note that the vertical scale differs by an order of magnitude from that used in Fig. 3.
A small but significant mismatch with observations is seen in both the two models.
Comparing with Fig. 3(c) and with corresponding sound-speed difference (Fig. 4a), we can say that the solar sound speed
is slightly bigger than in the first model (Fig. 5a) in the domain $0.85 R_\odot<r<0.9 R_\odot$ and slightly smaller in the domain
$0.8 R_\odot<r<0.85 R_\odot$ (we can say nothing about deeper layers because, as with model S, the residuals become distorted
by much bigger inaccuracy in the tachocline). The second model (Fig. 5b) shows better agreement in the interval $0.8 R_\odot<r<0.85 R_\odot$, but disagreement in the interval $0.85 R_\odot<r<0.9 R_\odot$ is made bigger. These two models were used as a reference
in the structural inversions for the adiabatic exponent $\Gamma_1$ in Paper I. The results are shown in Paper I by Figs 12(a) and 12(b)
for the first and the second model, respectively; they are in agreement with the conclusions drawn from the analysis of group velocities.
The magnitude of the mismatch between the models and the data is quite small: it calls for corrections of about $1\cdot 10^{-4}$ in the profile of the adiabatic exponent.
\begin{figure}
\begin{center}
\includegraphics[width=1.0\linewidth]{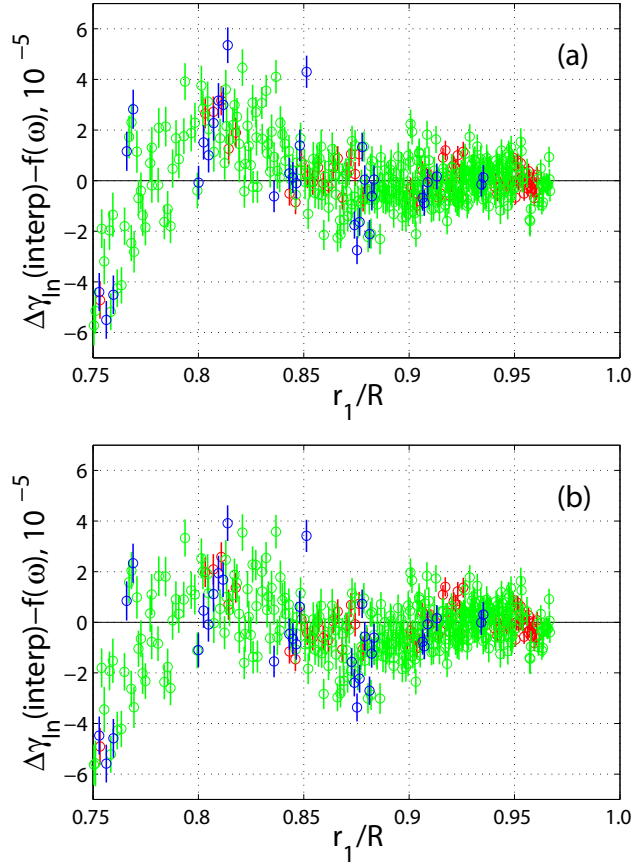}
\end{center}
\caption{As Fig. 3(c), but obtained with eigenfrequencies of two solar models with nearly-optimal parameters of the adiabatic
part of the convective envelope, constructed with SAHA-S3 equation of state and discussed in Paper I. (a): model with
$Y=0.250,\,\,Z=0.008$; (b) model with $Y=0.245,\,\,Z=0.010$.}
\label{f5}
\end{figure}

\section[]{Calibration of the envelope model}
We now implement the group-velocity measurements to the calibration of the the global parameters of the solar convective
envelope---specific entropy in the adiabatically-stratified layers and two parameters of the chemical composition.
As in Paper I, we compare with seismic data the 3-dimensional grids of envelope models calculated with four different versions
of the equation of state. The models are described in detail in Paper I.

Fig. 6 illustrates the potential possibility of simultaneous measurement of the parameters of the convective envelope by showing the response of the mismatch in $\gamma_{ln}$ between the Sun and the model to the variation of the specific entropy, helium abundance,
and heavy-element abundance in the model. Fig. 6(a) shows the residuals obtained with the best-fit model in the grid of models calculated
with SAHA-S3 equation of state. Figs 6(b) and 6(c) show the response of the residuals to the variation of the specific entropy (controlled by a mixing-length parameter $\alpha$) and to the variation of the helium abundance $Y$. Each of the two variations produces
a quasi-periodic signal in the residuals by changing the profile of the adiabatic exponent in the He II ionization region.
The two signals look similar, but closer inspection reveals that they have different phase: smaller entropy (bigger $\alpha$)
shifts the He II ionization to greater depths (it also shifts the location of the upper turning points,
but these two effects do not compensate for each other).
\begin{figure}
\begin{center}
\includegraphics[width=1.0\linewidth]{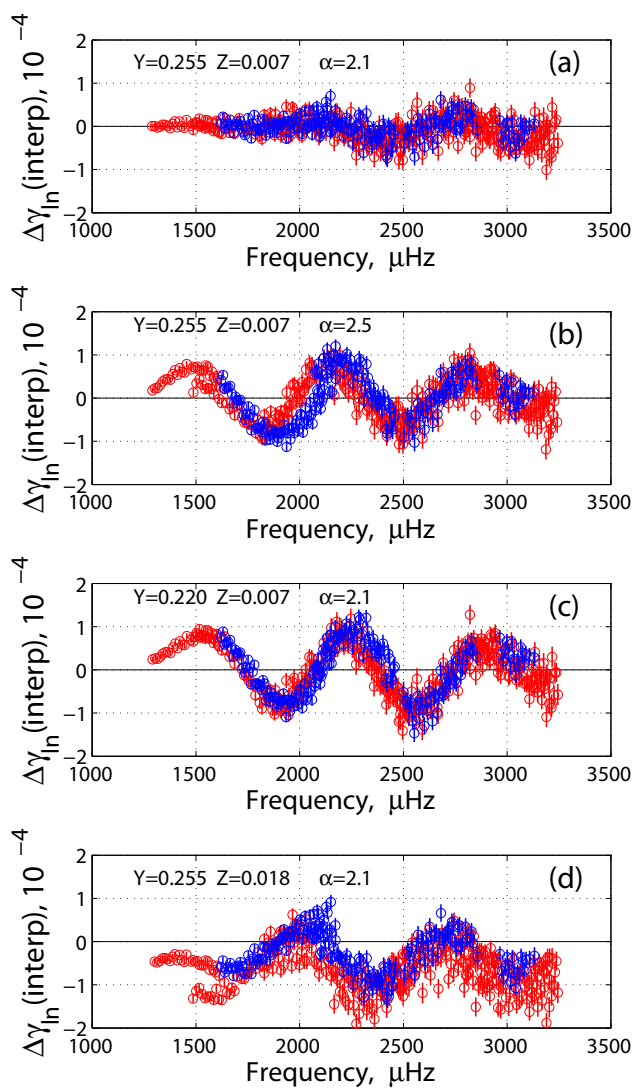}
\end{center}
\caption{Difference between solar values of $\gamma_{ln}$ and model predictions, for the ``best-fit'' model in the 3-D grid
of envelope models (a), for a model of the same chemical composition but with different specific entropy in the adiabatically-stratified
layers (b), model which differs from the best-fit model in He abundance Y (c), and model which differs in the heavy-element abundance Z (c). Red circles show the results in the $\tilde w$-range between 4000 s and 7000 s (lower turning points between
$0.85 R_\odot$ and $0.933 R_\odot$), blue circles---for $\tilde w>7000$ s.}
\label{f6}
\end{figure}
Principally, it is the availability of both the amplitude and phase of the He II ionization signal in the solar oscillation
frequencies which allows separate measurement of helium abundance and entropy; this property was used in the first seismic
measurements of the solar He abundance \citep[e. g.][]{Vorontsov91} \citep[for extended discussion, see][]{Vorontsov92}.
The signal produced by the variation of the heavy-element abundance (Fig. 6d) is more complicated; a distinctive feature
of this signal is that bigger $Z$ shifts $\Delta\gamma_{ln}$ to negative values, which signals that the sound speed in the model
is too small (smaller $Z$ brings smaller values to the adiabatic exponent $\Gamma_1$).

The results of the calibration are illustrated by Fig. 7. In these computations, the frequency range of the input data was limited by
lower frequencies (below 2 mHz), where the difference in absolute values of the observational and theoretical frequencies
is relatively small, and observational and theoretical values of $\gamma_{ln}$ were compared directly, without any interpolation
to common penetration depths (common values of $\tilde w$). The goodness of fit was measured by the merit function ($\chi^2$ per
degree of freedom)
\begin{equation}
M^2={1\over N}\sum\limits_{l,n}
  \left[{\gamma_{ln}^{\rm(obs)}-\gamma_{ln}^{\rm(model)}\over\delta\gamma_{ln}^{\rm obs}}\right]^2,
\end{equation}
where $\delta\gamma_{ln}^{\rm obs}$ is the $1\sigma$ uncertainty in the observational values of $\gamma_{ln}$ (defined by
equation 11) induced by the expected random errors in the oscillation frequencies. The solar p-mode frequencies were measured from the 1-yr SOHO MDI power spectra by the technique described in \citep[][]{Vorontsov09,Vorontsov13b} (data set 4 of Paper I). As in Paper I,
the input date were limited by $\tilde w>4000$ s (inner turning points $r_1>0.85 R_\odot$) to eliminate modes with theoretical
frequencies distorted by the inadequate description of the solar tachocline. (The same data were used in the results shown in Fig. 6).
\begin{figure}
\begin{center}
\includegraphics[width=0.95\linewidth]{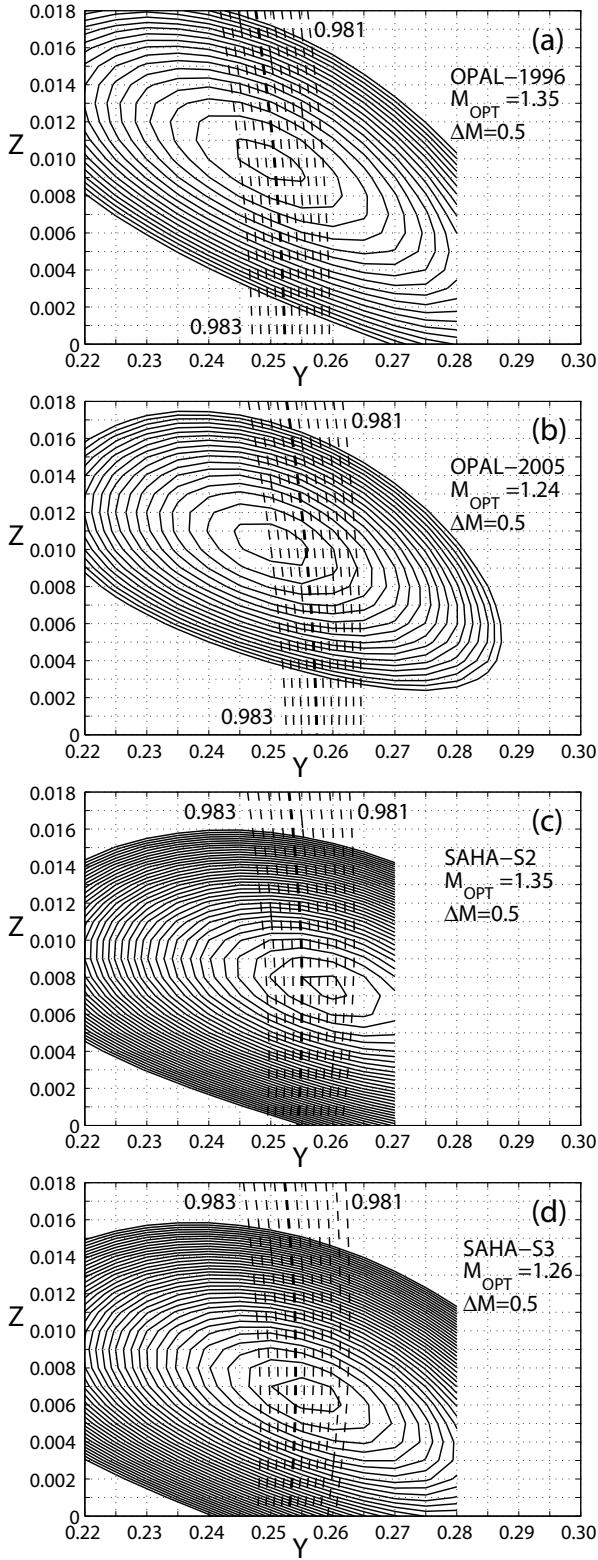}
\end{center}
\caption{Goodness of fit (merit function $M$) of envelope models calculated with (a) OPAL-1996, (b) OPAL-5005, (c) SAHA-S2 and
(d) SAHA-S3 equations of state. $M_{\rm opt}$ is the best value of the merit function, $\Delta M$ is the interval between
contour lines (solid curves). For each pair of $Y$ and $Z$, the specific entropy was choosen to optimize $M$. Dashed level lines
show the mass coordinate $m_{0.75}$ (equation 16) in the models optimized with respect to the specific entropy. The thick dashed
line is for $m_{0.75}=0.9822$.}
\label{f7}
\end{figure}

Dashed lines in Fig. 7 show the values of the dimensionless mass coordinate taken at $r=0.75 R_\odot$,
\begin{equation}
m_{0.75}=m(0.75 R_\odot)/M_\odot,
\end{equation}
for models which have optimal specific entropy. When helioseismic structural inversion is performed into the radiative interior,
this parameter determines the density profile obtained in the solar core. The ability of the inversion to fit low-degree
measurements depends on a proper value of this parameter (corresponding effects in the oscillation frequencies come from the
effects of gravity perturbation in the high-density core). Successful inversion into the deep interior requires
$m_{0.75}\simeq 0.9822$ (this finding is discussed in more detail in Paper I).

The maximum-likelihood values of $Y$ and $Z$, resulted from the calibration, are in agreement with those reported in Paper I;
they are in the range of $Y$=0.245--0.260 and $Z$=0.006--0.011.
On average, the optimal values of $Y$ are now slightly bigger (by about 0.005), and optimal values of $Z$ are slightly smaller
(by about 0.001).

The major difference with the results reported in Paper I is that different versions of the equation of state allow to achieve
nearly the same optimal values for the merit function. In our vision, the failure of our group-velocity analysis to distinguish between the performance of different equations of state comes principally from the more limited 
amount of the input data. The entire (and most valuable) p$_1$-ridge, for example, is only used for
evaluating the group velocities of $n=2$ waves (see equation 11). Also, evaluating $(\partial\omega/\partial n)_L$
using central differences over large frequency intervals (up to 1 mHz) brings an excessive averaging and loss of spatial resolution.

The attractive feature of the calibration with $\gamma_{ln}$ is its ultimate simplicity, which brings better confidence to the results.
Two functions of frequency were allowed in the calibration of Paper I to account for the uncertain near-surface effects; in general, this strategy makes the results more ambiguous. No allowance for any uncertainties was given in the calibration which is described above.
Another convenient feature is that $\gamma_{ln}$, as dimensionless quantity, is invariant to the homology rescaling of the hydrostatic model; the calibration reported in Paper I had to implement the rescaling as an extra (fourth) parameter, to allow small corrections to the solar radius.

Since the near-surface uncertainties are not eliminated completely in the analysis, we performed several numerical experiments
to address the stability of the calibration. Fig. 8(a) shows shows the result obtained with using interpolated model values of $\gamma_{ln}$ (see section 3). Fig. 8(b) shows the result obtained when the grid of models was recalculated using
\citet[][]{Canuto91} convection theory as an alternative to standard prescription. This modification brings the absolute values of the theoretical frequencies to somewhat better agreement with observations. In Fig. 8(c) we address the result of an artificial experiment
where outer boundary conditions in the eigenfrequency computations (here, we implemented ``zero'' boundary conditions, with setting
Lagrangian pressure perturbation to zero) were shifted from the temperature minimum to the photospheric level. This modification
makes the discrepancy between observational and theoretical frequencies bigger. Fig. 7(d) shows the result obtained when the observational frequency set was replaced with earlier measurements \citep[][data set 3 of Paper I]{Schou99}. We conclude from these experiments that calibration is moderately stable, but the accuracy of measuring the chemical-composition parameters of the solar envelope is not better
than 0.005 in $Y$ and 0.002 in $Z$.
\begin{figure}
\begin{center}
\includegraphics[width=0.95\linewidth]{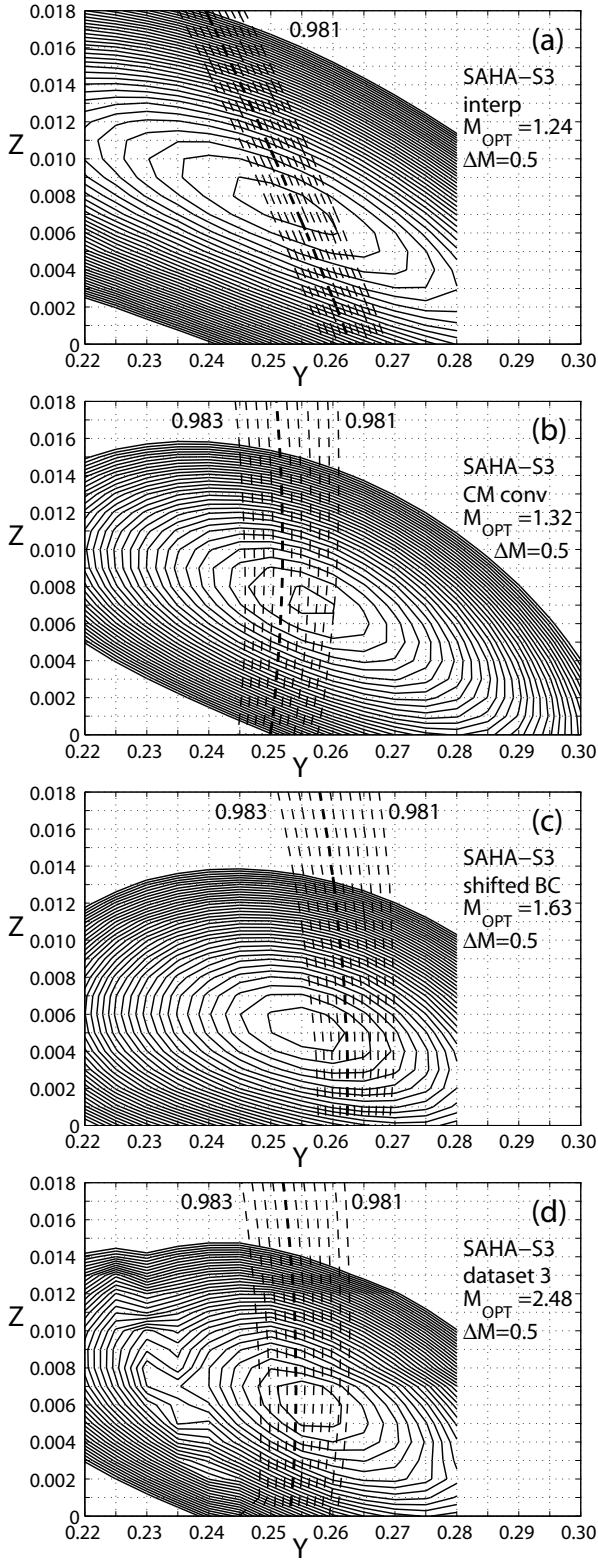}
\end{center}
\caption{Stability of the calibration of envelope models calculated with SAHA-S3 equation of state (each panel has to be compared with Fig. 7d). (a) the result obtained when observational and theoretical values of $\gamma_{ln}$ were compared with using interpolation
to common values of $\tilde w$; (b) the result obtained with models calculated using an alternative prescription of the convection
theory; (c) effect of changing the outer boundary conditions in eigenfrequency computations; (d) effect of changing the observational data set (see text).}
\label{f8}
\end{figure}

\section[]{Discussion}
The analysis of the solar p-mode data in terms of group velocities of surface waves represents an alternative tool of helioseismic
measurements, which can be used productively for validating the results obtained with more traditional methods of solar seismology.
The distinctive feature of the group velocities is their enhanced sensitivity to the solar stratification in the bottom part
of the acoustic cavity, and smaller sensitivity to the uncertain effects of the near-photospheric layers, when compared with oscillation frequencies. This property allows to make the technique of seismic analysis more simple and transparent.

Calibration of the chemical-composition parameters $Y$ and $Z$ in the solar convective envelope confirms our previous results (Paper I),
obtained with using a much more sophisticated analysis of the solar oscillation frequencies. All our results support strongly the downward revision of heavy-element abundances reported in recent spectroscopic measurements \citep[][]{Asplund09}.

In further work, the development of inversion techniques implementing the concept of group velocity may be a significant step forward.
Another issue is related with raw data analysis at high degree $l$, where accurate frequency measurements represent a very difficult task
\citep[for an account of the current efforts, see][]{korz13}. An interesting approach may consist in measuring the group velocity directly, as a slope of the p-mode ridge in the $l-\nu$ power spectra.

\section*{Acknowledgments}
In this work, V.A.B. and S.V.A. were supported by the RBRF grant 12-02-00135-a.

\appendix
\section{The variational principle}

In operator form, the equations of linear adiabatic oscillations of a spherically-symmetric star can be written as
\begin{equation}
\rho_0\omega^2{\bf u}=H_0{\bf u},
\end{equation}
where ${\bf u}$ is the displacement field, and linear integro-differential operator $H_0$ is defined as
\begin{equation}
H_0{\bf u}=\nabla p'+\rho'\nabla\psi_0+\rho_0\nabla\psi',
\end{equation}
\begin{equation}
p'=-\rho_0c^2\nabla\cdot{\bf u}+\rho_0{\bf u}\cdot\nabla\psi_0,
\end{equation}
\begin{equation}
\rho'=-\nabla\cdot\left(\rho_0{\bf u}\right),
\end{equation}
\begin{equation}
\nabla^2\psi'=4\pi G\rho',
\end{equation}
where $\psi$ is gravitational potential, subscript zero designates equilibrium values of corresponding physical quantities,
and their Eulerian perturbations are designated by prime.

We take the scalar product of the both sides of equation (A1) with ${\bf u}^*$, were star stands for complex conjugate,
and integrate over volume $V_R$ occupied by the star. Using Gauss theorem, it is straightforward to show that
\begin{eqnarray}
&&\omega^2\int\limits_{V_R}\rho_0{\bf u}^*\cdot{\bf u}\,dv
=\int\limits_{V_R}\Bigg[{1\over\rho_0c^2}p'^*p'+\rho_0N^2u_r^*u_r\nonumber\\
&+&\rho_0\left({\bf u}^*\cdot\nabla\psi'+{\bf u}\cdot\nabla\psi'^*\right)
+{1\over 4\pi G}\nabla\psi'^*\cdot\nabla\psi'\Bigg]dv\nonumber\\
&+&\int\limits_{S_R}\Bigg[u_r^*p'-{1\over 4\pi G}\psi'^*\left({\partial\psi'\over\partial r}+4\pi G\rho_0u_r\right)\Bigg]ds,\nonumber\\
\end{eqnarray}
where $S_R$ is unperturbed spherical outer boundary, $u_r$ is radial component of ${\bf u}$, and $N$ is Brunt-V\"ais\"al\"a frequency,
\begin{equation}
N^2=-g_0\left({d\ln\rho_0\over dr}+{g_0\over c^2}\right),
\end{equation}
where $g_0$ is unperturbed gravitational acceleration.

We now define a quadratic functional
\begin{equation}
\Phi=\int\limits_{V_R}\!\left({\bf u}^*\cdot H_0{\bf u}-\omega^2\rho_0{\bf u}^*\cdot{\bf u}\right)dv,
\end{equation}
and reduce the right-hand of this expression to the integral in radial coordinate using separation of spatial variables
specified by equations (1, 4, 6). We transform the surface term using standard boundary condition for gravity perturbations
\begin{equation}
\left[{dP\over dr}-4\pi G\rho_0U+{l+1\over r}P\right]_{r=R}=0
\end{equation}
which physical meaning is the continuity of the gravitational potential and its gradient on the deformed solar surface.
The result is
\begin{equation}
\Phi=\int\limits_0^R{\cal L}\,dr+R^2\left[Up_1+{l+1\over 4\pi Gr}P^2\right]_{r=R}
\end{equation}
with
\begin{eqnarray}
&&{\cal L}=r^2\Bigg\{{1\over\rho_0c^2}p_1^2+\rho_0N^2U^2
-2\rho_0\left(U{dP\over dr}+{L\over r}WP\right)\nonumber\\
&+&{1\over 4\pi G}\left[\left({dP\over dr}\right)^2+{L^2\over r^2}P^2\right]
-\rho_0\omega^2\left(U^2+W^2\right)\Bigg\},
\end{eqnarray}
where
\begin{equation}
W=LV
\end{equation}
and
\begin{equation}
p_1=-\rho_0c^2{dU\over dr}+\left(\rho_0g_0-{2\rho_0c^2\over r}\right)U+{\rho_0c^2\over r}LW.
\end{equation}
We now consider $\Phi$ as a homogeneous quadratic function of ``fields'' $U, W, P$ and their derivatives, which depend on
$\omega,\,l$ and structural variables of the equilibrium model as ``parameters'' (note that the Eulerian pressure perturbation
$p_1$ is not a ``field'' but an auxiliary variable). It can be verified directly that ${\cal L}$, defined by the equation (A11),
satisfies the Euler-Lagrange equations
\begin{equation}
{d\over dr}{\partial{\cal L}\over\partial\dot{U}}-{\partial{\cal L}\over\partial U}=0,
\end{equation}
\begin{equation}
{d\over dr}{\partial{\cal L}\over\partial\dot{W}}-{\partial{\cal L}\over\partial W}=0,
\end{equation}
\begin{equation}
{d\over dr}{\partial{\cal L}\over\partial\dot{P}}-{\partial{\cal L}\over\partial P}=0,
\end{equation}
where dot designates the radial derivative (equation A14 is equivalent to the radial component of the momentum equation,
equation A15---to the horizontal component of the momentum equation, and equation A16---to the Poisson's equation for gravity perturbations).

We now designate as $\delta_U$ the first variation of a corresponding quantity induced by a small variation of $U$ with
keeping the two other ``fields'' $W$ and $P$ and all the ``parameters'' unchanged. In a similar way, we introduce variations
$\delta_W,\,\delta_P,\,\delta_L$, and $\delta_\omega$. Using integration by parts and equations (A14--A16),we have
\begin{equation}
\delta_U\int\limits_0^R{\cal L}\,dr
=\int\limits_0^R\left[{\partial{\cal L}\over\partial U}\delta U+{\partial{\cal L}\over\partial\dot{U}}{d\over dr}(\delta U)\right]dr
=\left[{\partial{\cal L}\over\partial\dot{U}}\delta U\right]_0^R
\end{equation}
and similar expressions for $\delta_W\int_0^R{\cal L}\,dr$ and $\delta_P\int_0^R{\cal L}\,dr$. For variations of $\Phi$ we obtain,
using equations (A10, A11),
\begin{equation}
\delta_U\Phi=R^2\left[U\delta_Up_1-p_1\delta U\right]_{r=R},
\end{equation}
\begin{equation}
\delta_W\Phi=R^2\left[U\delta_Wp_1\right]_{r=R},
\end{equation}
\begin{equation}
\delta_P\Phi=0,
\end{equation}
\begin{eqnarray}
\delta_L\Phi&=&\delta(L^2)\int\limits_0^R\left(\omega^2\rho_0r^2V^2+{1\over 4\pi G}P^2\right)dr\nonumber\\
&+&R^2\left[U\delta_Lp_1+{\delta l\over 4\pi Gr}P^2\right]_{r=R},
\end{eqnarray}
\begin{equation}
\delta_\omega\Phi=-\delta(\omega^2)\int\limits_0^R\rho_0r^2\left(U^2+W^2\right)dr,
\end{equation}
where boundary condition for gravity perturbations (equation A9) was used in deriving equation (A20).
Due to the definition of $\Phi$ (equation A8), its variations sum to zero; since we do not change the parameters
of the equilibrium model, we have
\begin{equation}
\delta\Phi=\left(\delta_U+\delta_W+\delta_P+\delta_L+\delta_\omega\right)\Phi=0.
\end{equation}
Using equations (A18-A22), we get
\begin{eqnarray}
&&\delta(L^2)\int\limits_0^R\left(\omega^2\rho_0r^2V^2+{1\over 4\pi G}P^2\right)dr\nonumber\\
&-&\delta(\omega^2)\int\limits_0^R\rho_0r^2\left(U^2+W^2\right)dr\nonumber\\
&+&R^2\left[U\delta p_1-p_1\delta U\right]_{r=R}+{\delta l\over 4\pi G}RP^2(R)=0,
\end{eqnarray}
where $\delta p_1$ is the net variation of $p_1$. We will now assume that the (homogeneous and conservative) mechanical outer boundary
condition can be written in a form
\begin{equation}
AU+Bp_1=0,
\end{equation}
where $A$ and $B$ do not depend on $l$ and $\omega$ (an example is the so-called ``zero'' boundary condition,
$\nabla\cdot{\bf u}=0$). The variations $\delta U$ and $\delta p_1$ are then related as
$U\delta p_1=p_1\delta U$, the third term in the equation (A24) vanishes, and we arrive to
\begin{eqnarray}
&&\delta(\omega^2)\int\limits_0^R\rho_0r^2\left(U^2+L^2V^2\right)dr\\
&=&\delta(L^2)\int\limits_0^R\left(\omega^2\rho_0r^2V^2+{1\over 4\pi G}P^2\right)dr
+{\delta l\over 4\pi G}RP^2(R),\nonumber
\end{eqnarray}
the equation which relates small variations of frequency $\omega$ and degree $l$.
In the outer space ($r>R$), $P(r)$ is a regular solution to the Laplace equation,
$P(r)\propto r^{-l-1}$, and we have
\begin{eqnarray}
\int\limits_R^\infty P^2dr&=&-\int\limits_R^\infty{r\over l+1}P{dP\over dr}dr\nonumber\\
&=&{1\over 2}{R\over l+1}P^2(R)+{1\over 2(l+1)}\int\limits_R^\infty P^2 dr,
\end{eqnarray}
using integration by parts, which gives
\begin{equation}
RP^2(R)=(2l+1)\int\limits_R^\infty P^2dr.
\end{equation}
Using $(2l+1)\delta l=\delta(L^2)$, an alternative form of the equation (A26) is thus
\begin{eqnarray}
&&\delta(\omega^2)\int\limits_0^R\rho_0r^2\left(U^2+L^2V^2\right)dr\nonumber\\
&=&\delta(L^2)\left[\omega^2\int\limits_0^R\rho_0r^2V^2dr+{1\over 4\pi G}\int\limits_0^\infty P^2dr\right],
\end{eqnarray}
which gives equation (3) of section 2. We note that similar derivation was addressed, using somewhat different approach,
in \citep{Vorontsov06}. Due to inaccuracy in the treatment of the effects of gravity perturbations, term with gravity perturbations
(second term in the right-hand side of the equation A29) has been lost in the final result of \citep{Vorontsov06}.

We now address the derivation of the expression (9) for $(\partial\omega/\partial n)_L$.
In the analysis which is described above, we replace volume $V_R$ occupied by the star by a smaller volume $V_b$ bounded
by a spherical surface $S_b$ of radius $r=r_b$. We choose $r_b$ somewhere in the domain where the wave propagation is close to
that of purely acoustic waves, in the low-density envelope where gravity perturbation $\psi'$ is described by a solution
to the Laplace equation, which is regular at $r=\infty$. The boundary condition on gravity perturbations (equation A9)
which was implemented at $r=R$ is also satisfied at $r=r_b$, as well as everywhere in between (term with $\rho_0$ in equation A9 is small,
and was retained in the derivation of $(\partial\omega/\partial L)_n$ only to make it more general. When working with
solar oscillations, gravity perturbations in the outer envelope may be discarded at any degree $l$ due to low density). Keeping $L$ constant, the equation (A24) is replaced with
\begin{equation}
\delta(\omega^2)\int\limits_0^{r_b}\rho_0r^2\left(U_i^2+W_i^2\right)dr
=r_b^2\left[U_i\delta p_{1,i}-p_{1,i}\delta U_i\right]_{r=r_b},
\end{equation}
where we use subscript $i$ to designate solutions in the domain $r<r_b$ (solutions which satisfy central boundary conditions).
We now consider solutions in the external domain $r_b\le r\le R$, which satisfy surface boundary condition specified by the
equation (A25), with frequency-independent $A$ and $B$. In the similar way, we get
\begin{equation}
\delta(\omega^2)\int\limits_{r_b}^R\rho_0r^2\left(U_e^2+W_e^2\right)dr
=-r_b^2\left[U_e\delta p_{1,e}-p_{1,e}\delta U_e\right]_{r=r_b},
\end{equation}
where subscript $e$ designates the external solutions. At resonant frequencies, the internal and external solutions match
each other. Adding the equations (A31, A32), we have
\begin{equation}
r_b^2\left[U_e p_{1,i}-p_{1,e}U_i\right]_{r=r_b}
=\delta(\omega^2)\int\limits_0^R\rho_0r^2\left(U^2+W^2\right)dr.
\end{equation}
The radial-displacement function $U(r)$ and Eulerian pressure perturbation $p_1(r)$ are related by the differential equation
\begin{equation}
{dp_1\over dr}=\left(\omega^2-N^2\right)\rho_0U-{g_0\over c^2}p_1,
\end{equation}
which comes from the radial component of the momentum equation (A2) in Cowling approximation.
Using equations (5) and (8), which specify the wave function $\psi_p(\tau)$, the Wronskian of the internal and external solutions
in the left-hand side of the equation (A32) can be represented in terms of $\psi_p(\tau)$, and we have
\begin{equation}
{1\over\omega^2}\left(\psi_{p,i}{d\psi_{p,e}\over d\tau}-\psi_{p,e}{d\psi_{p,i}\over d\tau}\right)
=\delta(\omega^2)\int\limits_0^R\rho_0r^2\left(U^2+W^2\right)dr.
\end{equation}
Let $\theta$ is the radial phase integral, which takes values of $\pi(n+1)$ at resonant frequencies,
and $\delta\theta=\pi\delta n$ is variation of this phase integral induced by a small deviation of frequency $\omega$
from a resonant frequency. Representing $\psi_{p,i}$ and $\psi_{p,e}$ in the vicinity of the matching point by harmonic functions
of $\omega\tau$ with a small phase difference $\delta\theta$, we get
\begin{equation}
\pi\delta n={\omega\over\psi_p^2+{1\over\omega^2}\left({d\psi_p\over d\tau}\right)^2}
\delta(\omega^2)\int\limits_0^R\rho_0r^2\left(U^2+L^2V^2\right)dr,
\end{equation}
where $\psi_p$ is the wave function at the resonant frequency, and we arrive to the equation (9) of section 2.
Note that the derivation only assumes the applicability of the high-frequency approximation (harmonicity of $\psi_p(\tau)$)
in the vicinity of the matching point, not in the entire acoustic cavity.

\bsp
\label{lastpage}
\end{document}